\documentclass[aps,prb,twocolumn,amsfonts,amssymb,showpacs,eqsecnum]{revtex4}
\usepackage{amsmath}
\usepackage{amssymb}
\usepackage{graphicx}

\begin{document}

\title{Kinetics of geminate recombination of subdiffusing particles
in the presence of interparticle interaction}
\author{A. I. Shushin}
\affiliation{Institute of Chemical Physics, Russian Academy of Sciences, 117977, GSP-1,
Kosygin str. 4, Moscow, Russia}

\begin{abstract}
The kinetics of geminate subdiffusion-assisted reactions (SDARs) of
interacting particles is analyzed in detail with the use of the
non-Markovian fractional Smoluchowki equation (FSE). It is suggested
that the interparticle interaction potential is of the shape of
potential well and reactivity is located within the well. The
reaction kinetics is studied in the limit of deep well, in which the
FSE can be solved analytically. This solution enables one to obtain
the kinetics in a simple analytical form. The analytical expression
shows that the SDAR kinetics fairly substantially depends on the
mechanism of reactivity within the well. Specific features of the
kinetics are thoroughly analyzed in two models of reactivity: the
subdiffusion assisted activated rate model and the first order
reaction model. The theory developed is applied to the
interpretation of experimental kinetics of photoluminescence decay
in amorphous $a$-Si:H semiconductors governed by geminate
recombination of electrons and holes which are recently found to
undergo subdiffusive relative motion. Analysis of results
demonstrates that the subdiffusion assisted activated rate mechanism
of reaction is closer to reality as applied to amorphous $a$-Si:H
semiconductors. Comparison of experimental and theoretical kinetics
allowed for obtaining some kinetic parameters of the systems under
study: the rate of escaping from the well and the parameter
characterizing the deviation of the subdiffusive motion from the
conventional one.
\end{abstract}

\pacs{05.40.Fb, 02.50.-r, 76.20.+q}
\maketitle

\bigskip

\section{Introduction}

Diffusion assisted reactions (DARs) is the important stage of a
large number of chemical and physical condensed phase
processes.\cite{Ri,Ha,Ca} In many of these processes the DARs are
known to be strongly affected by interaction between reacting
particles. The effect of the interaction on the DAR kinetics is
actively studied both experimentally and theoretically for a long
time.\cite{Ri,Ca}

In the case of conventional diffusion the problem of the analysis of
DAR kinetics reduces to solving the Smoluchowski equation for the
pair distribution function (PDF) of reacting particles. This
equation is fairly complicated and can, in general, be solved only
numerically. As for analytical study, usually it is made with the
use of steady state analytical solutions.\cite{Ri,Ca} The general
time dependent analytical solutions can be found only for very few
interaction potentials, for example, in the case Coulomb
interaction.\cite{Noo} This solutions, however, are very cumbersome
and complicated for applications.

Some years ago much more simple and rigorous method of analytical
solution of the Smoluchowski equation was proposed, which is
applicable in the practically interesting limit of deep well of
attractive interaction (reactivity assumed to be localized in the
well).\cite{Shu1,Shu2,Shu3} The solution shows that the interaction
strongly manifests itself in the reaction kinetics resulting in the
long life time of particles within the well (i.e. caging). The time
evolution of the PDF of pairs captured and reacting in the well
appears to be non-exponential.\cite{Shu2,Shu3} This specific feature
of the PDF time evolution shows itself, for example, in
non-exponential kinetics of geminate DARs with the long time tail of
inverse power type.\cite{Shu2,Shu3}

In this work we will consider the kinetics of geminate reaction of
interacting particles undergoing subdiffusive motion. Recall that
subdiffusion is a certain type of anomalous diffusion, which is
characterized by the anomalously slow time dependence of the mean
square of displacement $\langle r^2 (t) \rangle \sim t^{\alpha}$
with $\alpha < 1$.\cite{Blum,Met1}  Recently, the specific features
of the kinetics of subdiffusion assisted reactions (SDARs) is a
subject of active
discussions.\cite{Yuste,ShushinPRE,Shushin,ShushinNJ,Henry,Vlad,Fed,
Sung,Seki,Yuste2004,Sokolov06} The anomaly of diffusion is shown to
affect fairly strongly the reaction kinetics leading to the
effective slowing down of the reaction at long times, to the strong
fluctuations of concentrations of reacting particles at long times,
etc.

In the absence of interparticle interaction (i.e. in the case of
free subdiffusion) the time evolution of the PDF of subdiffusing
particles is usually described by the analog of diffusion
equation,\cite{Kenkre,Met1} which is called fractional diffusion
equation and in which the effect of diffusion anomaly shows itself
in anomalously long time memory. The predictions of the theory based
on fractional diffusion equation are analyzed in a large number of
papers (see, for example, reviews [8] and [9]).

As for the SDARs of interacting particles, these precesses are not
studied theoretically yet. The kinetics of them is also determined
by the corresponding PDF, but the PDF evolution is described by the
fractional Smoluchowski equation.\cite{Met2,Shushin} Similarly to
the case of conventional diffusion, the factional equation is much
more complicated for numerical and analytical analysis than that for
free diffusion. In this work we propose the analytical solution of
the fractional Smoluchowski equation in the limit of deep well,
assuming the reactivity to be localized within the well. With the
use of the obtained solution the PDF evolution and the kinetics of
geminate SDARs are analyzed in detail in this limit.

The analysis shows that, unlike the case of conventional DAR, the
SDAR kinetics strongly depends on the mechanism of reaction. In
particular, the kinetics appears to be essentially different for two
models of reactivity: the subdiffusion assisted activated rate model
[or, more generally, the kinetically (i.e. mobility) controlled
reaction model] and the first order reaction model. This strong
difference enables one to select the realistic reaction mechanism by
comparison of theoretical predictions with the experimental data.

The obtained results are applied to the interpretation of the
experimental kinetics of photoluminescence decay in amorphous
semiconductors $a$-Si:H resulting from geminate recombination of
photoexcited electrons ($e$) and holes
($h$).\cite{Street,Street1,SekiPRB} Electrons in these
semiconductors are known to be highly mobile, undergoing
subdiffusive (dispersive) migration, while holes are nearly
immobile. \cite{Street} Recently, fairly detailed experimental
investigation of the kinetics of geminate $e$-$h$ recombination at
different temperatures has been carried out\cite{SekiPRB} and
experimental results have been semiquantitatively described within
the free subdiffusion model. It is worth noting, however, that the
kinetics of the process under study is, clearly, significantly
affected by the Coulomb $e$-$h$ interaction which is quite strong in
the investigated semiconductors: for dielectric constant
$\varepsilon \approx 10$ characteristic for these semiconductors and
temperatures $T < 300 K$ the Onsager radius (distance, at which the
Coulomb interaction is equal to the thermal energy) is estimated as
$l_e > 50A $.

Analysis of theoretical kinetic dependences, obtained in this work,
shows that the activated rate model describes the experimental
results better than the first order reaction one. This analysis
allowed us to obtain the characteristic model parameters of the
system: the parameter $\alpha$ characterizing the diffusion anomaly,
the rate of escaping from the $e$-$h$ interaction well, etc., which
result in the best fitting, and estimated their dependence on
temperature. The proposed interpretation is compared with another
one applied in ref. [24] to treat some earlier experimental results
on the same process.

\section{Diffusion assisted reactions}

We start our analysis with the discussion of geminate reactions of
interacting particles assisted by conventional diffusion. The
reacting particles are assumed to be spherically symmetric. We also
assume that the interparticle interaction potential is  spherically
symmetric: $u({\bf r}) \equiv u(r) = U( r)/k_BT $, where $r = |{\bf
r}|$ is the interparticle distance, and is of the type of potential
well with the reaction barrier at short (contact) distance $r = d$,
the bottom at $r = r_b$ (see Fig. 1). It is suggested, in addition,
that the well is deep enough so that the activation energies of
escaping from the well and reaction are large: $u_a = - u(r_b) \gg
1$ and $u(d)-u(r_b) \gg 1$. In this limit the characteristic time of
reaction in the well is much larger than the time of equilibration
in the well and the long distance (Coulomb) part of the potential
can be characterized by the effective Onsager radius
\begin{equation} \label{gen00}
l_e = \left[\mbox{$\int_{r_b}^{\infty}$} \, dr \,
r^{-2}e^{u(r)}\right]^{-1}.
\end{equation}
nearly independent of $r_b$ [this radius satisfies the relation
$u(l_e) \approx 1$]. Noteworthy is that in the deep well limit the
explicit shape of the well of $u({r})$ at short distance turns out
to be not important for the kinetics: it can be, for example, of
parabolic shape with $r_b > d$ or edge type one with $r_b \approx d$
(see Fig. 1).

\begin{figure}
\includegraphics[height=10.0truecm,width=8.5truecm]{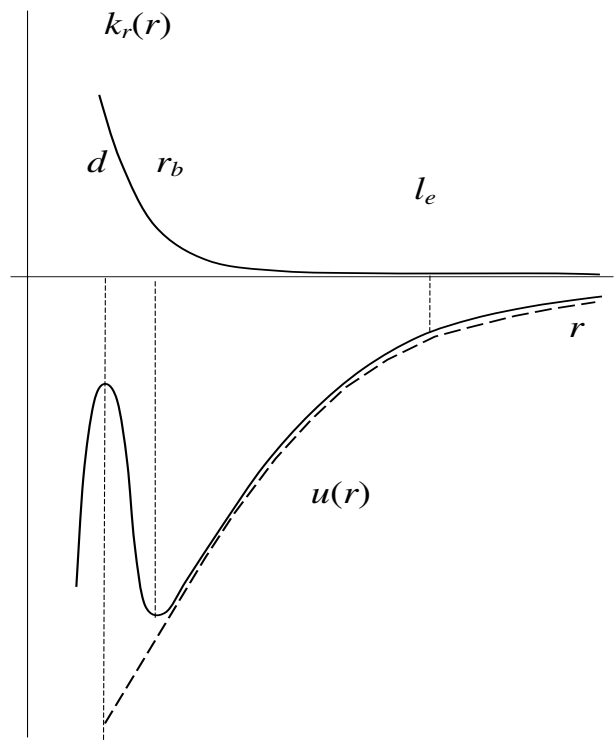}
\caption{Schematic picture of the distance dependence of the
interaction potential $u(r) = U(r)/(k_B T)$ and reactivity $k_r
(r)$. In this picture $d$ is the distance of closest approach, $r_b$
is the coordinate of the bottom, and $l_e$ is the Onsager radius.
The dashed line shows the Coulomb potential which leads to the
cusp-like well at $r=d$.}
\end{figure}

In the Markovian approach implied by the conventional diffusion
approximations the DAR kinetics is described by the PDF $\rho(r,t)$
of reactive pairs, satisfying the kinetic equation
\begin{equation} \label{gen01}
\dot \rho ({\bf r},t)= -\hat {\cal L} \rho({\bf r},t),
\end{equation}
where $\hat {\cal L}$ is the operator which determines space/time
evolution of the system under study. The form of this operator
depends on the process considered (see Sec. III).

In this work the observable under study is the geminate DAR
kinetics, i.e. the time dependent probability $Y_r(t)$ of geminate
reaction, which we will call the DAR yield. In our further analysis,
however, it will be more convenient to analyze the DAR flux
\begin{equation} \label{gen01d}
J_r (t) = \dot Y_r(t).
\end{equation}

To obtain the expression for DAR yield and DAR flux one needs to
specify the model of reactivity. In our work we assume that the
reactivity is localized within the well and consider two most well
known models of reactivity: the diffusion assisted activated rate
and first order reaction models. They correspond to two different
reaction mechanisms: kinetically controlled (controlled by relative
migration) and first order reaction controlled reactivity,
respectively. Possible examples of these reaction models are
discussed below as applied to analysis of experimental results (see
Sec. IV). Here we will restrict ourselves to discussion of the
mathematical details of the models.

1) {\it Activated rate model.} This model treats the DAR as a
diffusive flux over a barrier located at the reaction surface [in
the studied model of spherically symmetric particles this is the
barrier at $r = d$ (see Fig. 1)]. Some well known formulas for the
reaction yield obtained in this model are presented below (in Sec.
IIB). The activated rate model is, actually, a particular example of
the general class of kinetically controlled reaction models, which
predict the reaction rate proportional to the diffusion coefficient,
or more generally speaking, to the mobility of particles (Sec. IV).

2) {\it First order reaction model.} In this model the DAR flux
$J_r(t)$ is determined by the fluctuating rate $k_r[r(t)]$ of first
order reaction and is expressed in terms of the functional:
\begin{equation} \label{gen01a}
J_r(t) = \left\langle \! k_r [r(t)]\exp \bigg(\!\!\!-\int_0^t d\tau
\, k_r [r(\tau)]\bigg)\!\right\rangle_{\!r(t)},
\end{equation}
in which the average is made over the stochastic fluctuations of
$r(t)$ governed by process (\ref{gen01}).

In the considered approach the DAR flux can be written in a
universal form in terms of the the Green's function $\rho ({\bf
r},{\bf r}_i|t)$ of the stochastic Liouville equation
\begin{equation} \label{gen01b}
\dot \rho = -[\hat {\cal L} + k_r (r)]\rho + \delta (t)\delta ({\bf
r}-{\bf r}_i):
\end{equation}
\begin{equation} \label{gen01c}
J_r(t) =  - \int_{r>d} \! d{\bf r} \! \int \! d {\bf r}_i \,\dot
\rho ({\bf r},{\bf r}_i|t)\rho_i ({\bf r}_i),
\end{equation}
where $\rho_i ({\bf r})$ is the initial spatial distribution of
particles [naturally, in the activated rate model $k_r(r) = 0$].

\subsection{Mechanisms of stochastic motion}

Here we will briefly discuss some useful models for description of
relative jump-like stochastic motion of reacting particles and
analyze the validity of the diffusion approximation for description
of the process under study.

One of the most general models of spatial jump-like evolution of the
system is based on the Kolmogorov-Feller approach in which
\begin{equation} \label{gen02}
\hat {\cal L} = - w(1 -\hat {\cal P}),
\end{equation}
where $w$ is the jump rate, for simplicity assumed to be independent
of the coordinate, and $\hat P$ is the matrix of transition
probabilities satisfying the detail balance relation and the
normalization condition which in the coordinate representation
${\cal P}({\bf r},{\bf r}_i)$ for $\hat {\cal P}$ is written as
\begin{equation} \label{gen03}
\mbox{$\int$} d{\bf r}\, {\cal P}({\bf r},{\bf r}_i) = 1.
\end{equation}

In general, there is a large variety of jump models for $\hat {\cal
P}$ satisfying above relations. Here we  will discuss a class of
models especially convenient for theoretical analysis. These models
are based on the representation of the matrix $\hat {\cal P}$ in
terms of the differential Smoluchowski-like diffusion operator
\begin{equation} \label{gen04}
\hat L \rho = -D \nabla_{\bf r} (\nabla_{\bf r} \rho +
\rho\nabla_{\bf r} u),
\end{equation}
in which $\nabla_{\bf r}$ is the gradient operator in $\{{\bf
r}\}$-space and $D$ is the diffusion coefficient,\cite{Shu4,Shu5}
\begin{equation} \label{gen05}
\hat {\cal P} = \widetilde{P}(\hat L) = \int_0^{\infty} \! d\tau
\,e^{-{\hat L}\tau} P(\tau).
\end{equation}
In the representation (\ref{gen05}) the function $P(\tau)$ is
properly normalized: $\int_0^{\infty} d\tau \, P(\tau) = 1$, so that
$\hat {\cal P}$ can be considered as the operator of diffusive
evolution averaged over some distribution function $P(\tau)$ of
evolution times. Evidently, the operator $\hat {\cal P}$
(\ref{gen05}), which has the form of the Laplace transform of
$P(\tau)$, satisfies the normalization relation (\ref{gen03}).

The model (\ref{gen05}) is very useful for the analysis of
applicability of the diffusion (Smoluchowski) approximation which
appears to be valid in a wide region of parameters of the model and
times. The validity criterion can easily be obtained with the use of
eq. (\ref{gen01}) written for the Laplace transform in time
$\widetilde{\rho} (r,\epsilon ) = \int_0^{\infty} \!d\tau \,\rho
(r,\tau)e^{-\epsilon\tau}:$
\begin{equation}\label{gen06}
\epsilon \widetilde{\rho} =  -w (1-\hat {\cal P})\widetilde{\rho}.
\end{equation}
According to eq. (\ref{gen06}) for small $\epsilon/w \ll 1$, when
the left hand side of this equation is small, the operator $1-\hat
{\cal P}$ in the right hand side can be expanded in $\hat L\tau$:
$1-\hat {\cal P} \approx -\hat L {\bar t}$, where
\begin{equation}\label{gen07}
\bar t = \mbox{$\int_0^{\infty} dt \,t P(t)$}
\end{equation}
is the average time of diffusion-like evolution resulting in the
jump operator $\hat {\cal P}$. The effect of next orders of
expansion of $\hat {\cal P}$ in $\hat L\tau$ is evidently of higher
orders in $\epsilon/w \ll 1$ and therefore is negligibly small. The
correctness of this statement can also be demonstrated by expanding
the solution of eq. (\ref{gen06}) in the (complete) basis of
eigenfunctions of the operator $\hat L$.

The presented analysis shows that at relatively long times $t > 1/w$
the kinetics of processes, governed by jump-like operator
(\ref{gen05}), is quite accurately described by the corresponding
Smoluchowski equation.

In what follows we will restrict ourselves to this long time limit
of the reaction kinetics and correspondingly to the Smoluchowski
approximation. In addition, for simplicity, we will consider the
spherically symmetric geminate processes which are described by
distribution functions depending only on distance $r = |{\bf r}|$.

\subsection{Equations of diffusion approximation and two-state model}

In the considered case of spherically symmetric geminate reactions
the PDF $\rho (r,t)$, depending on the interparticle distance $r =
|{\bf r}|$, satisfies the Smoluchowki kinetic equation, which for
\begin{equation} \label{gen0}
\sigma (r,t)= r\rho (r,t)
\end{equation}
is written as
\begin{equation} \label{gen1}
\dot \sigma =  D \nabla_r (\nabla_r \sigma + \sigma\nabla_r
u)-k_r\sigma,
\end{equation}
where $\nabla_r = \partial /\partial r$  and $D$ is the diffusion
coefficient.

The function $\sigma (r, t)$ (\ref{gen1}) obeys the boundary
conditions $\sigma(r \rightarrow \infty,t) \rightarrow 0$ and
$\sigma(r \rightarrow 0,t) \rightarrow 0$, and the isotropic initial
condition
\begin{equation}\label{gen1b}
\sigma_i (r) = \sigma(r,0)= \delta(r-r_i)/(4\pi r_i)
\end{equation}
with $r_i \sim r_b$ corresponding to the creation of particles
within the well. This equation can also be represented in terms of
the Laplace transform $\widetilde{\sigma} (r,\epsilon ) = \int_0^t
\!d\tau \,\sigma (r,\tau)e^{-\epsilon\tau}$:
\begin{equation} \label{gen1d}
\epsilon \widetilde{\sigma} - \sigma_i=  D \nabla_r(\nabla_r
\widetilde{\sigma} + \widetilde{\sigma}\nabla_r
u)-k_r\widetilde{\sigma}.
\end{equation}
This representation appears to be more convenient for our further
analysis.

In general, eqs. (\ref{gen1}) and (\ref{gen1d}) cannot be solved
analytically. In the limit of deep well, however, the solution can
be obtained in a simple form by expansion in a small parameter
$\tau_r /\tau_e \ll 1$, where $\tau_r \sim l_e^2/D$ is the time of
equilibration within the well and $\tau_e \sim \tau_r e^{-u_a}$ is
the time of escaping from the well.\cite{Shu2,Shu3}

Analysis of this solution shows\cite{Shu3} that in the lowest order
in the parameter $\tau_r /\tau_e$ the Smoluchowski approximation
(\ref{gen1}) and (\ref{gen1d}) is equivalent to the model of two
kinetically coupled states. This two-state model treats the process
under study as transitions between the state within the well ($d < r
< l_e$), whose population is
\begin{equation} \label{gen1e}
n(t) = 4\pi \int_d^{l_e} \! dr \,r \sigma (r,t),
\end{equation}
and the state of free diffusion outside the well ($r > l$) described
by the distribution function $C(r,t) = r^{-1}c(r,t)$. The functions
$n(t)$ and $c(r,t)$ satisfy simple kinetic
equations,\cite{Shu2,Shu3} which can be written in the most compact
form in terms of Laplace transforms $\widetilde{n}(\epsilon)$ and
$\widetilde{c}(r,\epsilon)$:
\begin{subequations} \label{gen2}
\begin{eqnarray}
\epsilon\widetilde{n} &=&1+ [S_l^{-1}K_{+}
\widetilde{c}(l_e,\epsilon) - (K_- + w_r)\widetilde{n}] \qquad
 \label{gen2a}\\
\epsilon\widetilde{c} &=& [D \nabla_r^2 \widetilde{c} +
(S_lK_-\widetilde{n} - K_+\widetilde{c}) \delta (r-l_e)],\qquad
\label{gen2b}
\end{eqnarray}
\end{subequations}
where $S_l = (4\pi l_e)^{-1}$. The terms proportional to $K_{\pm}$
describe the above-mentioned kinetic coupling (transitions) between
the state within the well, located at $r = l_e $, and the free
diffusion state outside the well. In the considered limit $\tau_r
/\tau_e \ll 1$ the transition rates $K_{\pm}$ satisfy the
relations:\cite{Shu3}
\begin{equation} \label{gen3} K_{\pm}
\rightarrow \infty \quad \mbox{and} \quad K_{+}/K_{-} = K_e = Z_w,
\end{equation}
where
\begin{equation}\label{gen3a}
Z_w = \int_{d<r<l_e}dr \, r^2 e^{-u(r)}
\end{equation}
is the partition function for the well.

Equations (\ref{gen2}) are written for the initial condition
\begin{equation} \label{gen4}
n(0) = 1 \quad \mbox{and} \quad c(r,0) = 0.
\end{equation}
corresponding to the initial population of the well implied by eq.
(\ref{gen1b}). As to the boundary conditions for $c(r,t)$, they are
given by $l_e \nabla_r c (r,t) - c(r,t)|_{r=l_e} = 0$ and $c(r
\rightarrow \infty) = 0$.

The term $w_r \widetilde{n}$ in eq. (\ref{gen2a}) describes the
effect of the first order reaction in the well. In the considered
(Markovian) diffusion approximation the two above-mentioned models
of reactivity in the well result in the similar kinetic equations of
the form (\ref{gen2}). The only difference consists in the
analytical expression for $w_r$:

\paragraph{Activated reaction model $(k_r = 0)$.}
In the diffusion assisted activated reaction model, implying
activated diffusive passing over the barrier at $r \sim d$, one
gets\cite{Shu1,Shu2,Shu3}
\begin{equation} \label{gen3b3}
w_r = w_{r_a} =  \frac{D}{Z_w} \left(\int_{r\sim d} \! dr \,
r^{-2}e^{u(r)}\right)^{-1} .
\end{equation}
As we have already mentioned above this model is an example of a
large class of kinetically controlled reaction models, in which the
reaction rate is determined by the mobility of particles: $w_r \sim
D$. Some examples of such models are discussed in Sec. IV.

\paragraph{First order reaction model $(k_r \neq 0)$.}
In the model describing the reaction as a first order process with
rate $k_r (r)$ the expression for , for example, in quite realistic
case of relatively small values of $k_r$ within the
well\cite{Shu1,Shu2,Shu3}
\begin{equation} \label{gen3b1}
w_r = w_{r_f} = \langle k_r \rangle =  \frac{1}{Z_w} \int_{w}dr \,
k_r (r) r^2 e^{-u(r)}.
\end{equation}
In this equation the parameter $w$, used as a limit of integration,
denotes integration over the region near the bottom of the well ($d
< r < l_e$).

In both models the Markovian DAR kinetics, expressed in terms of the
Laplace transform $\widetilde{J}_r (\epsilon)$ of the DAR flux
(\ref{gen01d}), is proportional to the the Laplace transform
$\widetilde{n} (\epsilon)$ of the well population:
\begin{equation} \label{gen4a}
\widetilde{J}_r (\epsilon) = w_r \widetilde{n} (\epsilon),
\end{equation}
i.e to analyze the reaction kinetics one should find the time
dependent well population $n (t)$.

For our further discussion it is convenient to represent equations
(\ref{gen2}) in a matrix form:
\begin{equation} \label{gen2ab}
\epsilon \widetilde{\bf R} = -(\hat \Lambda + \hat W_r)
\widetilde{\bf R}+\widetilde{\bf R}_i , \;\;\mbox{where}\;\;
\widetilde{\bf R} = (\widetilde{n}, \widetilde{c})^{\top},
\end{equation}
$\widetilde{\bf R}_i = (1,0)^{\top}$ is the vector representation of
the initial condition (\ref{gen4}),
\begin{equation}
\hat \Lambda = \!\!\left[
\begin{array}{cc} \label{gen2ac}
K_{-} &  -S_l^{-1}K_{+}\int\! dr \delta(r\!-l_e) \\
-S_lK_{-}\delta(r-l_e)   &   -D \nabla_r^2 + K_{+} \delta (r\!-l_e)
\end{array}
\!\right]
\end{equation}
and
\begin{equation} \label{gen2ad}
\hat W_r = \!\!\left[
\begin{array}{cc}
w_{r} &  0 \\
0  &  0
\end{array}
\!\right].
\end{equation}

This representation shows that equations (\ref{gen2}) can be
considered as a two-state analog of the SLE (\ref{gen01b}),
corresponding to the two-state model of $k_r$-fluctuations in the
functional (\ref{gen01c}). Naturally, in this SLE the effect of
reactivity is represented by $(2\times 2)$- matrix of the form
(\ref{gen2ad}).

It is also worth emphasizing that the two-state model (\ref{gen2})
is actually based on the approximate replacement of the Smoluchowski
operator $\hat L_r = D \nabla_r (\nabla_r  + \nabla_r u)$ by the
simpler one $\hat \Lambda + \hat W_r$, operating on the reduced PDF
$\widetilde{\bf R}$. The approximation is valid in the limit of deep
well, when all eigenvalues of $\hat L_r$ representing population
relaxation within the well are much larger than the lowest one,
which describes quasistationary escaping from the well and reaction.
In the deep well limit the effect of high eigenvalues is negligibly
small and the kinetics of the process is quite accurately treated
within the approximation taking into account the coupling of the
lowest state in the well with the continuum of states outside the
well (which is just equivalent to the proposed two-state
model\cite{Shu3}). This main idea of the proposed method is
important point in our further analysis of SDARs.

It is also important to note that in the deep well limit the
two-state model is valid independently of the shape of the well near
the bottom of the well (see Fig. 1). The shape manifests itself only
in the value of partition function $Z_w$ defined in eq.
(\ref{gen3a}).

\subsection{Reaction kinetics}

In accordance with the relation (\ref{gen4a}) the DAR kinetics [i.e.
the reaction flux $J_r (t)$] is determined by that of the well
depopulation $n(t)$ which can be obtained by solution of eqs.
(\ref{gen2}) and subsequent inverse Laplace
transformation:\cite{Shu2,Shu3}
\begin{equation}\label{gen5}
n(t) = \frac{1}{2\pi i} \int\limits_{-i\infty+0}^{i\infty+0} \!\!
d\epsilon \frac{\exp (\epsilon t)}{\epsilon + w_r + l_e^2K_e
V(\epsilon)}.
\end{equation}

The function $V(\epsilon)$ is directly related to the Green's
function of the operator which controls diffusion outside the well
[with the reflective boundary condition $(l_e \nabla g - g)|_{r=l_e}
= 0$]\cite{Shu2,Shu3}
\begin{equation} \label{gen6} g(r,r_i|\epsilon) := \langle r |
\left(\epsilon + D\nabla^2_r \right)^{- 1} |r_i \rangle :
\end{equation}
\begin{equation} \label{gen7}
V(\epsilon) := 1/g(l_e,l_e|\epsilon) = D [l_{e}^{-1} +
(\epsilon/D)^{1/2}].
\end{equation}

Substitution of the expression (\ref{gen7}) into eq. (\ref{gen5})
leads to the following formula for the well population
$n(t)$:\cite{Shu3}
\begin{eqnarray}
n(t)&=& \frac{1}{2\pi i} \int_{-i\infty+0}^{i\infty+0} d\varepsilon
\frac{\exp[\varepsilon (w_{_0} t)]}{1 + \varepsilon +
\gamma \varepsilon^{1/2} }\nonumber\\
&=& \frac{\varepsilon_{+} {\Phi}_1^{+}(w_{_0} t)-\varepsilon_{-}
{\Phi}_1^{-}(w_{_0} t)}{\varepsilon_+ -
\varepsilon_-},\qquad\qquad\quad \label{gen12}
\end{eqnarray}
where \, $\gamma = \sqrt{(w_e/w_{_0})(l_e^2w_e/D)}$, \,
$\varepsilon_{\pm} = \mbox{$\frac{1}{2}$}\gamma \pm i \sqrt{1-
\frac{1}{4}\gamma^2}$ are the roots of equation $z^2 - \gamma z + 1
= 0$, \, and \,
\begin{eqnarray}
\Phi_1^{\pm}(z) = [1-{\rm erf}(\varepsilon_{\pm}\sqrt{z})]\exp
(\varepsilon_{\pm}^{2}z) \label{gen13a}.
\end{eqnarray}
The rate
\begin{equation} \label{gen14}
w_{_0} = w_e + w_r
\end{equation}
is a sum of the rate of escaping from the well
\begin{equation} \label{gen15}
w_e =  \frac{D}{Z_w} \left(\int_{r_b}^{\infty} \! dr \,
r^{-2}e^{u(r)}\right)^{-1} = Dl_e/Z_w
\end{equation}
and the rate of reaction in the well $w_r$ [see eqs. (\ref{gen3b3})
and (\ref{gen3b1})].

Specific features of the well population $n(t)$ (\ref{gen12}) are
analyzed in detail in refs. [6] and [7]. In general, this function
is non-exponential and the analytical properties are essentially
determined by the parameter $\gamma$ introduced in eq.
(\ref{gen12}). The physical meaning of this parameter is clear from
the relation $\gamma = \sqrt{(w_e/w_{_0})(l_e^2w_e/D)} \sim
l_e/l_D$, in which $l_D = \sqrt{D/w_0}$ is the average distance of
diffusive motion during the life time $\tau_0 = w_0^{-1}$ of the
particle in the well. This relation shows that if the well is deep
enough,  $l_D \gg l_e$ and therefore $\gamma \ll 1$. The parameter
$\gamma$ controls the qualitative change of the analytical behavior
of the kinetics:
\begin{eqnarray}
n(t) &=& \exp (-w_{_0} t) \;\; \mbox{at} \; \tau
\lesssim \ln (1/\gamma),\label{gen16a}\\
n(t) &\sim & 1/t^{3/2} \qquad \;\; \mbox{at} \; \tau \gg \ln
(1/\gamma).\label{gen16b}
\end{eqnarray}
In the limit of small $\gamma \ll 1$ more detailed analysis of the
dependence $n(t)$ is possible,\cite{Shu2,Shu3} however, here we will
restrict ourselves to these simple relations only.

In addition to the kinetic time dependences the steady state
characteristics of type of the total DAR yield $Y_r^{\infty} =
Y_r(t\to\infty)$ are also of certain interest for further
applications. They can easily be obtained with the use of
expressions derived. For example,
\begin{equation} \label{gen9}
Y_r^{\infty} = \int_0^{\infty} dt \, J_r(t) = w_r\widetilde{n}(0)
=\frac{w_r}{w_r + w_e}.
\end{equation}

\section{Subdiffusion assisted reactions}

\subsection{Equations in subdiffusion approximation}

In the case of subdiffusive motion the evolution of $\rho (r,t)$ is
described the subdiffusion (fractional) variant of the Smoluchowski
equation.\cite{Met1} This equation can be derived within the
continuous time random walk approach\cite{Met1} for jump-like motion
of particles assuming long time tailed behavior of the probability
density function of waiting times for jumps and using formula
(\ref{gen05}) for the distribution of jumps ${\cal P}({\bf r},{\bf
r}_i)$. Similar to the case of the conventional diffusion considered
(Sec. IIA) the fractional Smoluchowski equation is obtained in the
limit of relatively weak deviation of $\rho({\bf r},t)$ from the
equilibrium PDF, when $ (\overline{t}\hat L)\rho \ll \rho$ and
therefore one can expand $\hat {\cal P}=\widetilde{\hat L}$ in
powers of $\hat L\overline{t}$:
\begin{equation} \label{sub1}
\dot \sigma =  D_s {}_{0}^{} D_t^{1-\alpha} \nabla_r (\nabla_r
\sigma + \sigma\nabla_r u)],
\end{equation}
where $\nabla_r = \partial /\partial r$ , $D_s$ is the subdiffusion
coefficient, and
\begin{equation} \label{sub2}
{}_{0}^{}D_t^{1-\alpha} \sigma(r,t) = \frac{1}{\Gamma
(\alpha)}\frac{\partial}{\partial t}\int_0^t
\frac{dt'}{(t-t')^{1-\alpha}}\sigma(r,t')
\end{equation}
is the Riemann-Liouville fractional derivative\cite{Samko} with
$\alpha < 1$. The function $\sigma (r, t)$ (\ref{sub1}) satisfies
the boundary and initial conditions similar to those used for the
case of conventional diffusion  [see Sec. IIB].

The corresponding equations for the Laplace transform
$\widetilde{\sigma} (r,\epsilon )$ is written as
\begin{equation} \label{sub4a}
\epsilon \widetilde{\sigma} - \sigma_i=  D_s \epsilon^{1-\alpha}
\nabla_r(\nabla_r \widetilde{\sigma} + \widetilde{\sigma}\nabla_r
u).
\end{equation}

It is seen that from mathematic point of view in the absence of
reaction the difference of the subdiffusion variant of the
Smoluchowski equation for Laplace transform $\widetilde{\sigma}$
from the conventional Markovian equation reduces to the replacement
of $D$ with $D_s \epsilon^{1-\alpha}$. The problem becomes more
complicated, however, when one is going to analyze the effect of
reactivity. Below we will obtain the kinetic equations describing
the SDAR processes in the two models of reactivity proposed above:
the activated rate model and the model of first order reaction.

Since the coordinate part of the subdiffusion differential equations
(\ref{sub1}) and (\ref{sub4a}) coincides with that of the
Smoluchowski equation (\ref{gen1}) and (\ref{gen1d}) the two state
model is still applicable though with some modification of the form
of time dependences or, as applied to equations for Laplace
transforms, the dependences on the parameter $\epsilon$ [i.e. the
functions $\widetilde{n}(\epsilon)$ and
$\widetilde{c}(r,\epsilon)$].

\subsection{Activated rate reaction in the well}

In the model of subdiffusion assisted activated rate reaction in the
well the corresponding fractional analog of the kinetic equation
(\ref{gen2ab}) for $\widetilde{\bf R} = (\widetilde{n},
\widetilde{c})^{\top}$ can be written by taking into account that
the fractional Smoluchowski equations (\ref{sub2}) and (\ref{sub4a})
differ from the conventional ones only in operators which determine
the time dependence of the PDF. As for the coordinate operators,
they are similar in both equations differing only in diffusion
coefficients: $\hat L_{r_s} = D_s \nabla_r (\nabla_r + \nabla_r u)
\sim \hat L_r = D\nabla_r (\nabla_r + \nabla_r u)$. This fact is
important because the two-state model actually reduces to the
special two-state representation of the coordinate operator.
Correspondingly, if in the deep well limit this model is valid in
the case of conventional diffusion it is also valid as applied to
the subdiffusion processes since the validity conditions (emphasized
in the end of Sec. IIB) appeal only to the characteristic properties
of the coordinate operator (eigenvalues, describing population
relaxation in the well, should be mach larger than the lowest
eigenvalue representing the escape rate). The similarity of
coordinate operators: $\hat L_{r_s}= (D_s/D)\hat L_{r}$, enables us
to easily obtain the corresponding subdiffusion variant of the
two-state representation of $\hat L_{r_s}$: $ \hat \Lambda_s + \hat
W_{r_s}^{\alpha} = (D_s/D)(\hat \Lambda + \hat W_r)$, in which we
set $\hat W_{r_s} = [(D_s/D)\hat W_r]^{1/\alpha}$ to reveal the
anomalous dimensionality of $D_s$. Thus, finally, we arrive at
equation
\begin{equation}
\label{sub5} [\epsilon + \epsilon^{1-\alpha} (\hat \Lambda_s + \hat
W_{r_s}^{\alpha})] \widetilde{\bf R}=\widetilde{\bf R}_i,
\;\mbox{where}\;\, \widetilde{\bf R} = (\widetilde{n},
\widetilde{c})^{\top}\!,
\end{equation}
$\widetilde{\bf R}_i = (1,0)^{\top}$ is the initial condition [see
eq. (\ref{gen2ab})],
\begin{equation}
\hat \Lambda_s = \!\left[
\begin{array}{cc} \label{sub5a}
K^{\alpha}_{-} &  -S_l^{-1}K^{\alpha}_{+}\int\! dr \delta(r\!-l_e) \\
-S_lK^{\alpha}_{-}\delta(r\!-l_e)   &   -D_s \nabla_r^2 +
K^{\alpha}_{+} \delta (r\!-l_e)
\end{array}
\!\right]
\end{equation}
and
\begin{equation} \label{sub5b}
\hat W_{r_s} = \!\!\left[
\begin{array}{cc}
w_{r_s} &  0 \\
0  &  0
\end{array}
\!\right].
\end{equation}
is the matrix of SDAR rates in the well.

The parameters in these equations are similar to those in eqs.
(\ref{gen2}) [see eqs. (\ref{gen3}) and (\ref{gen3a})], except the
effective rate $\bar w_{r_s}$ of (anomalous) reaction in the well
modeled by subdiffusive passing over the barrier at $r \sim d$:
\begin{equation} \label{sub6b}
w_{r_s} = \left[\frac{D_s}{Z_w} \left(\int_{r\sim d} \! dr \,
r^{-2}e^{u(r)}\right)^{-1} \right]^{1/{\alpha}}.
\end{equation}
Recall that in our calculation we take the limit (\ref{gen3}), i.e.
$K^{\alpha}_{\pm} \rightarrow \infty $ with $ K_e =
K^{\alpha}_{+}/K^{\alpha}_{-}  = Z_w$.

Solution of eqs. (\ref{sub5}) and inverse Laplace transformation
give for the well population
\begin{equation}\label{sub7}
n(t) = \frac{1}{2\pi i} \int\limits_{-i\infty+0}^{i\infty+0} \!\!
d\epsilon \frac{\exp (\epsilon t)}{\epsilon +\epsilon^{1-\alpha}
[w_{r_s}^{\alpha} + l_e^2K_e V(\epsilon)]}.
\end{equation}

As in the case of conventional diffusion the function $V(\epsilon)$
is determined by the Green's function of the operator describing
subdiffusive relative motion of particles outside the well [with the
reflective boundary condition $(l_e \nabla g - g)|_{r=l_e} = 0$]
\begin{equation} \label{sub8}
g(r,r_i|\epsilon) = \langle r | \left(\epsilon^{\alpha} +
D_s\nabla^2_r \right)^{- 1} |r_i \rangle :
\end{equation}
\begin{equation} \label{sub9}
V(\epsilon) = 1/g(l_e,l_e|\epsilon) = D_s [l_{e}^{-1} +
(\epsilon^{\alpha}/D_s)^{1/2}].
\end{equation}

Formulas (\ref{sub7}) - (\ref{sub9}) reduce the problem of
calculating $n(t)$ to the evaluation of the Green's function
(\ref{sub8}) and subsequent inverse Laplace transformation
(\ref{sub7}):
\begin{eqnarray}
n(t)\!\!&=&\!\! \frac{1}{2\pi i} \int_{-i\infty+0}^{i\infty+0}
\frac{d\varepsilon}{\varepsilon^{1-\alpha}} \frac{\exp[\varepsilon
(w_{_{0_s}} t)]}{1 + \varepsilon^{\alpha} +
\gamma_a \varepsilon^{\alpha/2}}\qquad \nonumber\\
&=& \frac{\varepsilon_{+}^a\Phi_{\alpha}^{+} (w_{_{0_s}}
t)-\varepsilon_{-}^a\Phi_{\alpha}^{-} (w_{_{0_s}}
t)}{\varepsilon_{+}^a - \varepsilon_{-}^a},\qquad\; \label{sub10}
\end{eqnarray}
where 
\begin{equation} \label{sub10c}
\gamma_a = (l_e^2 w_{e_s}^{\alpha}/D_s)^{1/2}
(w_{e_s}/w_{_{0_s}})^{\alpha/2},
\end{equation}
the parameters $\varepsilon_{\pm}^a = \mbox{$\frac{1}{2}$}\gamma_a
\pm i \sqrt{1- \mbox{$\frac{1}{4}$}\gamma_a^2}$ are similar to those
introduced earlier [see eq. (\ref{gen12})], and $\Phi_{\alpha}^{\pm}
(\tau)$ are expressed in terms of the Mittag-Leffler function
$L_{\alpha/2} (-x)$:\cite{Erd}
\begin{eqnarray}\label{sub11}
\Phi_{\alpha}^{\pm} (t)& = &L_{\alpha/2}
[-\varepsilon_{_\pm}^a(w_{_{0_s}}t)^{\alpha}]\nonumber\\
 &=& \frac{1}{2\pi i}
\int_{-i\infty+0}^{i\infty+0}\!d\varepsilon \,
\frac{\exp[\varepsilon (w_{_{0_s}}t)]}{\varepsilon +
\varepsilon_{_\pm}^a \varepsilon^{1-\alpha/2} } .
\end{eqnarray}
It is easily seen that for $\alpha = 1$ $\Phi_{\alpha}^{\pm} (t)$
coincide with functions defined in eq. (\ref{gen13a}). Similarly to
formulas for conventional diffusion in eq. (\ref{sub10}) we have
introduced the rate
\begin{equation} \label{sub12}
w_{_{0_s}} = \left(w_{e_s}^{\alpha} +
w_{r_s}^{\alpha}\right)^{1/\alpha}
\end{equation}
in which
\begin{equation} \label{sub12a}
w_{e_s} = (D_s l_e/Z_w)^{1/\alpha}
\end{equation}
is the rate of subdiffusive escaping from the well and $w_{r_s}$ is
the anomalous SDAR rate in the well [see eqs. (\ref{sub6b})].

Unlike the conventional diffusive well depopulation kinetics
(\ref{gen12}), the subdiffusive kinetics (\ref{sub10}) is
non-exponential at all times. Moreover, in the most interesting case
of small $\gamma \ll 1$ with high accuracy one can neglect the last
term in denominator of the integrand in eq. (\ref{sub10}) and get
the expression
\begin{equation}\label{sub13}
n (t) \approx L_{\alpha}[-(w_{_{0_s}}\!t)^{\alpha}],
\end{equation}
which predicts the behavior $n (t) \sim (w_{_{0_s}}\!t)^{-\alpha}$
at $w_{_{0_s}} \!t \gg 1$.

With the use of relation (\ref{gen4a}), which in the subdiffusion
case is written as
\begin{equation} \label{sub14}
\widetilde{J}_r (\epsilon) = \epsilon^{1-\alpha}{
w}_{r_a}^{\alpha}\widetilde{n}(\epsilon),
\end{equation}
formula similar to eq.  (\ref{sub7}) can also be obtained for the
SDAR flux $J_r (t)$
\begin{eqnarray}
J_r(t)&=& \frac{w_{_{0_s}}}{2\pi i}\,
\frac{{w}_{r_a}^{\alpha}}{w_{_{0_s}}^{\alpha}}
\int_{-i\infty+0}^{i\infty+0}d\varepsilon\, \frac{\exp[\varepsilon
(w_{_{0_s}} t)]}{1 +
\varepsilon^{\alpha} + \gamma_a \varepsilon^{\alpha/2} }\nonumber\\
&=&\frac{({w}_{r_a}/w_{_{0_s}})^{\alpha}}{\varepsilon^a_{_+} \!-
\varepsilon^a_{_-}} \frac{d}{dt}\!
\left[\varepsilon^a_{_-}\Phi_{\alpha}^{+} (t) - \varepsilon^a_{_+}
\Phi_{\alpha}^{-} (t)\right]. \qquad\label{sub15}
\end{eqnarray}

This expression shows the important peculiarity of the activated
variant of the SDAR kinetics: the time dependence of the SDAR flux
$J_r (t)$ is essentially different from that of the well population
$n(t)$. Unlike relations obtained above for the conventional
diffusion and prediction of the expression (\ref{sub13}), in the
considered limit $\gamma \ll 1$ at relatively short times $\tau
\lesssim 1/\gamma^{1/\alpha} $ the flux $J_r (t) \sim \dot
L_{\alpha}[-(w_{0_s}\! t)^{\alpha}]$ so that the time dependence of
the SDAR flux is of inverse power type:
\begin{eqnarray}
J_r (t) &\sim & (w_{0_s}\! t)^{-(1+\alpha)} \quad\; \mbox{for} \; 1<
w_{0_s}\! t
< \gamma^{-2/\alpha},\label{sub16a}\\
J_r(t) &\sim & (w_{0_s}\! t)^{-(1+\alpha/2)} \;\; \mbox{for} \;
w_{0_s}\! t
> \gamma^{-2/\alpha}.\qquad\quad \label{sub16b}
\end{eqnarray}

Interestingly the expressions for steady state characteristics like
the total SDAR yield $Y_{r_s}^{\infty} = Y_{r_s} (t \to \infty)$ are
fairly similar to those derived above (see Sec.II) for the case of
conventional DAR, though with $w_r$ and $w_e$ replaced by
$w_{r_s}^{\alpha}$ and $w_{e_s}^{\alpha}$, respectively. In
particular,
\begin{equation} \label{sub17}
Y_r^{\infty} = \int_0^{\infty} dt \, J_r(t) = \widetilde{J}_r(0) =
\frac{ w_{r_s}^{\alpha}}{w_{r_s}^{\alpha} + w_{e_s}^{\alpha}}.
\end{equation}

\subsection{First order reaction in the well}

In the case of first order reaction in the well the SDAR kinetics is
described by more complicated non-Markovian stochastic Liouville
equation\cite{Shu1,Shu2,Shu3} which is the non-Markovian analog of
eq. (\ref{gen1}). In what follows, however, for simplicity we will
restrict ourselves to the more simple two-state model whose
subdiffusive variant is represented by eqs. (\ref{sub5}). With the
use of the method recently proposed in ref. [7] the corresponding
two state non-Markovian kinetic (stochastic Liouville) equation for
the vector $\widetilde{\bf R} = (\widetilde{n},
\widetilde{c})^{\top}$ can be derived in the following
form:\cite{ShushinPRE,Shushin,ShushinNJ}
\begin{equation} \label{sub18}
(\hat \Omega + \hat \Lambda_{s} \hat
\Omega^{1-\alpha})\widetilde{\bf R} = {\bf R}_i, \;\; \mbox{where}
\;\;\, \widetilde{\bf R} = (\widetilde{n}, \widetilde{c})^{\top},
\end{equation}
$\widetilde{\bf R}_i = (1,0)^{\top}$ is the initial condition,
\begin{equation} \label{sub19}
\hat \Omega = \epsilon + \hat W_{r_f}, \;\;\mbox{and}\;\; \hat
W_{r_f} = \!\!\left[
\begin{array}{cc}
w_{r_f} &  0 \\
0  &  0
\end{array}
\!\right]
\end{equation}
with $w_{r_f}$ defined in eq. (\ref{gen3b1}).

Solution of this equation (in the limit $K^{\alpha}_{\pm}
\rightarrow \infty$ with $K^{\alpha}_{+}/K^{\alpha}_{-} = Z_w$)
yields for the population of the well
\begin{equation}\label{sub20}
n(t)= \frac{1}{2\pi i}\!
\int\limits_{-i\infty+\varepsilon_{\!_f}}^{i\infty+\varepsilon_{\!_f}}\!\!\!
\frac{d\varepsilon}{\varepsilon^{1-\alpha}}\,
\frac{\exp[(\varepsilon - \varepsilon_{\!_f})(w_{e_a}t)]}{1 +
\varepsilon^{\alpha} + \gamma_{_f}
(\varepsilon-\varepsilon_{\!_f})^{\alpha/2}}.
\end{equation}
where 
\begin{equation} \label{sub21}
\varepsilon_{\!_f} = w_{r_{\!f}}/w_{e_s} \;\; \mbox{and} \;\;
\gamma_{_f} = (l_e^2 w_{e_s}^{\alpha}/D_s)^{1/2}.
\end{equation}
Naturally, in the model of first order reaction the SDAR flux $J_r
(t)$ is proportional to the well population:
\begin{equation} \label{sub22}
J_r (t) = w_{r_{\!f}} n (t)
\end{equation}
similarly to the case of conventional diffusion.

Expressions (\ref{sub20}) and (\ref{sub22}) show that in the first
order reaction model the SDAR kinetics differs from that obtained in
the activated rate model [eqs (\ref{sub10}) and (\ref{sub15})]. In
particular, in the most interesting limit of small $\gamma_{_f} \ll
1$ and for weak reactivity $w_{r_{\!_f}} \ll w_{e_s}$ at short times
$t < 1/(\gamma_{_{f}}^{1/\alpha}w_{e_s} )$
\begin{equation} \label{sub23}
J_r (t) = w_{r_{\!_f}} n (t) \approx
w_{r_{\!_f}}e^{-w_{r_{\!f}}t}L_{\alpha}[-(w_{e_s}t)^{\alpha}].
\end{equation}
More detailed analysis allows one to conclude that for $\gamma_{_f}
\ll 1$ over a wide region of times the time dependence of the SDAR
flux is of inverse power type given by:
\begin{eqnarray}
J_r (t) &\sim &  t^{-\alpha} \qquad\;\;\, \mbox{for} \; 1< w_{e_s}\!
t < \gamma_{_f}^{-2/\alpha}, w_{e_s}/w_{r_f};\nonumber\\
J_r(t) &\sim & t^{-(1+\alpha/2)} \; \mbox{for} \;  w_{{e_s}}t \gg
w_{{e_s}}/w_{r_f},\,\gamma_{_f}^{-2/\alpha}.
\qquad\quad \nonumber
\end{eqnarray}
In other words, the long time behavior of $J_r (t)$ is similar in
both models of reactivity, however at intermediate times the first
order reaction model predicts slower kinetics than the activated
rate one.

It is of certain interest to note that despite this difference in
the SDAR kinetics the formula for the SDAR yield in the first order
reaction model appears to coincide with that in the the activated
rate model [see eq. (\ref{sub17})]: $Y_r^{\infty}  =
w_{r_f}^{\alpha}/(w_{r_f}^{\alpha} + w_{e_s}^{\alpha}).$

\section{Electron-hole recombination kinetics}

Here we will apply the obtained results to describing the kinetics
of photoluminescence decay in amorphous semiconductors $a$-Si:H
governed by subdiffusion assisted geminate recombination of
photogenerated charge carries: electrons ($e$) and holes
($h$).\cite{Street,Street1,SekiPRB} The geminated $e$-$h$
recombination is strongly influenced by the attractive interaction
$u(r)$ of $e$- and $h$-quasiparticles. At $e$-$h$ distances $r$ much
larger than the size $l_{eh}$ of these quasiparticles the potential
$u(r)$ is of the Coulomb form
\begin{equation}\label{app1}
u (r) \stackrel{r \gg l_{eh}}{=} -l_e/r,\;\; \mbox{with}\;\; l_e =
e^2/(\epsilon_s k_B T),
\end{equation}
where $l_e$ is the Onsager radius [see eq. (\ref{gen00})] and
$\epsilon_s \approx 10$ is the static dielectric constant for the
semiconductor $a$-Si:H.\cite{Street,SekiPRB} Note that $l_{eh}$ is
probably of order of the mean spacing in this semiconductor. Is is
also worth noting that at short distances $r \sim r_b \ll l_{eh}$
the potential is expected to be somewhat flattened because of the
effect of finite size of charge distribution at these distances
resulting in the finiteness of $e$-$h$-interaction at $r \ll
l_{eh}$. This effect can qualitatively be understood by considering
the potential of interaction between point-like charge with the
homogeneously changed sphere which is known to be of parabolic shape
and finite at the center of the sphere. In the case of quantum
particles the qualitatively same effect is also expected as one can
see form the analysis of the electronic terms of the simplest
molecules $H_2$ and $H_2^+$.

However, independently of behavior of the potential at short
distances this potential is of the shape of the potential well (see
Fig. 1) and in the deep well limit the theory proposed is valid
independently of the well shape, as it has been mentioned above. The
characteristic features of well shape manifest themselves only in
the value of the partition function $Z_w$ and thus in values of
rates $w_r$ and $w_e$ which are considered as adjustable parameters
anyway.

It is worth noting that the Onsager radius $l_e$ is fairly large for
the systems considered (with the dielectric constant $\epsilon
\approx 10$) for the temperatures of experiments, $ 100 < T < 170\,
({\rm K})$, we get $140 > l_e > 80\, ({\rm A})$. In such a case the
assumption that the geminate $e$-$h$ pairs are initially created
within the well, i.e. at $r < l_e$ looks quite reasonable. The large
value of the Onsager radius $l_{e} \gg r_b \sim l_{eh}$ implies the
large depth of the potential well. In addition, large value of $l_e$
also ensures the validity of the diffusion approximation for
description of the jump like motion of electrons applied in our
analysis.

The exact shape of the potential and the mechanisms of jump-like
motion at short distances $r \sim d < l_{eh}$, which, according to
the above discussion, determines the mechanism of reaction in the
well, can hardly been found explicitly. One of the goals of this
section is the selection of the proper model of recombination
process in the well, i.e., actually, selection of mechanism of
reactivity, by comparison of predictions of the models with
experimental results. In our analysis we will consider two of them
discussed above:

a) {\it  First order reaction model.} This model assumes that
$e$-$h$ recombination in the well is the first order reaction whose
rate is determined by the direct charge-transfer exchange
interaction\cite{Ri} [the model implies the exponential distance
dependence of the reaction rate $k_r (r) \sim \exp (- r/r_e)$]. This
model is traditionally applied to describing the kinetics of
condensed phase recombination.\cite{Ri}

{b) {\it  Activated rate model.} The activated rate model is based
on the mechanism of diffusive passing over a barrier which can be
considered as kinetically controlled reaction (i.e. reaction
controlled by mobility of particles). This model is quite realistic
as applied to charge recombination in polar liquids in which the
mean-force interaction potential of charges is known to have a
fairly high potential barrier at short distances $r \sim
d$.\cite{Ri,Nin} As for solids, one can hardly expect similar
barrier in them, in general. Note, however, that the activated
passing over the barrier is only one example of kinetically
controlled reaction processes. Another example is the reactions
limited by jump-like migration  with small jump rates at short
distances (naturally, one should take into account the discreteness
of space at short distances). In such a case, these jumps of small
rate can control the rate of reaction. The small value of jump rates
at short distances $r \sim d$ can result from large difference in
energy (energy gap) between the states of migrating electron and the
final electron state in the ion pair. Moreover, the rate of jumps
into this final reacting state (this jumps are probably
irreversible) can be small both for positive and negative energy
difference (of large absolute value).\cite{Ovch} In the case of
small values of rates of irreversible jumps into the final state the
kinetics of the processes is correctly described by the
Smoluchowski-like equation with partially reflective boundary
condition, in which the reflective term is determined the
above-mentioned small (reactive) rate of jump in the final state.
Analysis of this equation shows that the two-state representation of
this equation (for the Laplace transform $\widetilde{\bf R}$) is
also of the form (\ref{gen2ab}) and (\ref{sub5}) in the cases of
normal diffusion and subdiffusion, respectively. It is important to
emphasize that in the subdiffusion variant of the two-state model
(\ref{sub5}) the term, describing the kinetically controlled
reaction in the well, is represented in the form
$\epsilon^{1-\alpha} \hat W_{r_s}^{\alpha}$, i.e. contains the
factor $\epsilon^{1-\alpha}$ describing the long time memory effects
caused by anomalous migration. {\it In other words, the kinetic
equations (\ref{gen2ab}) and (\ref{sub5}) derived in the activated
rate model appear to be valid for a number of models of kinetically
controlled reactions in the well, some of which are quite applicable
to solid state recombination reactions}.

To apply these models one needs to specify the kinetic scheme of the
process under study which allows one to relate the calculated
kinetics functions with the observed dependences. The scheme was,
actually, implied in the analysis made in refs. [21] and [26]. We
assume that the $e$-$h$ recombination results in the formation of
the fluorescing product ($P$) whose population will be hereafter
denoted as $n_p (t)$. The luminescence intensity $I(t)$ can easily
be determined from the simple kinetic scheme:
\begin{equation}\label{app2}
e + h \stackrel{J_r (t)}{\longrightarrow} P \stackrel{w_I +
w_d}{\longrightarrow} P_0
\end{equation}
which takes into account formation of the product $P$ with the rate
$J_r (t)$ within the well, radiationless deactivation into the
ground state $P_0$ with the rate $w_d$ and fluorescence with the
rate $w_i$. In a quite realistic limit of fast fluorescence and
deactivation when $w_d + w_i \gg w_e, w_r$ the kinetic equation for
the population $n_P (t)$
\begin{equation}\label{app4}
\dot n_P = -(w_d + w_i)n_P + J_r(t),
\end{equation}
corresponding to the scheme (\ref{app2}), predicts
\begin{equation}\label{app5}
I(t) \sim w_i n_P(t) \sim  [w_i/(w_d + w_i)]J_r(t).
\end{equation}
This formula shows the time dependence of the observed luminescence
intensity $I(t)$ is proportional to that of the recombination (DAR)
flux discussed above and therefore gives the direct information on
the kinetics of geminate $e$-$h$ recombination.

\begin{figure}
\includegraphics[height=11.0truecm,width=8.5truecm]{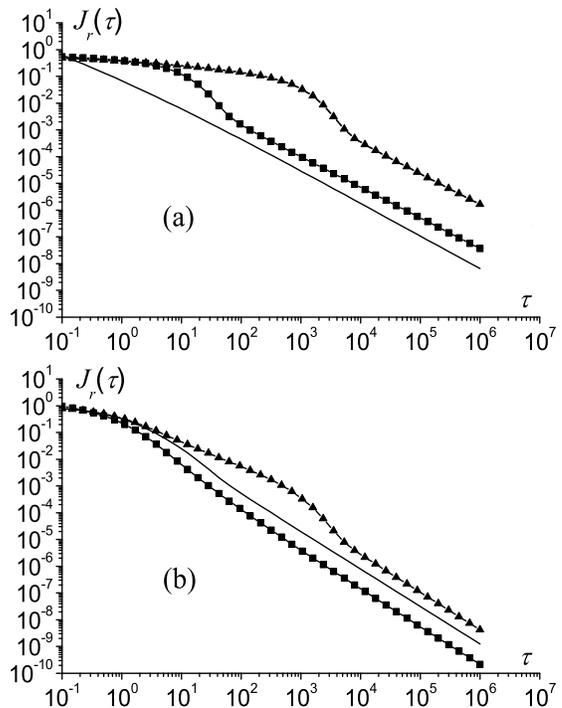}
\caption{The dependence of SDAR flux $J_r (\tau)$ (in arbitrary
units) on dimensionless time $\tau = w_{_{0_a}} t$ for two models of
reactivity in the well: the activated rate model (full lines) and
the first order reaction models (squares and triangles). The flux is
evaluated for: (a) $\alpha = 0.3$ and (b) $\alpha = 0.8$, as well as
for negligibly weak reactivity in the activated rate model and for
two values of reactivity in the first order reaction model:
$\varepsilon_f = w_{r_f}/w_{e_a} = 0.001$ (triangles) and
$\varepsilon_f = 0.1$ (squares). In evaluation the fixed value
$\gamma_a = \gamma_f = 0.2$ is used.}
\end{figure}

Noteworthy is that usually in the interpretation of experimental
results radiationless deactivation is not taken into consideration.
We mentioned this process only for the sake of completeness and
because typically it is faster than fluorescence transitions. It is
seen from the proposed kinetic scheme that the considered
experiments can not really give any information about relative
efficiency of these two processes and in what follows we are not
going to discuss this point.

Figure 2 demonstrates characteristic features of $J_r
(t)$-dependences obtained within the two models of reactivity in the
well discussed above. It is seen that at intermediate times the
first order reaction model predicts more sharp changing $J_r (t)$
than the activated rate model, though the asymptotic long-time
behavior of $J_r (t)$ is the same in both models in accordance with
derived analytical formulas.

The observed fairly substantial difference between predictions of
the two models allows for selecting the most realistic model by
comparison with experimental results. Close inspection shows that
the very smooth experimental dependences $J_r (t)$ are much closer
to those predicted by the activated rate model rather then by the
first order reaction one. This means that the activated rate model
seems to be closer to reality as applied to the process under study.

\begin{figure}
\includegraphics[height=8.0truecm,width=8.5truecm]{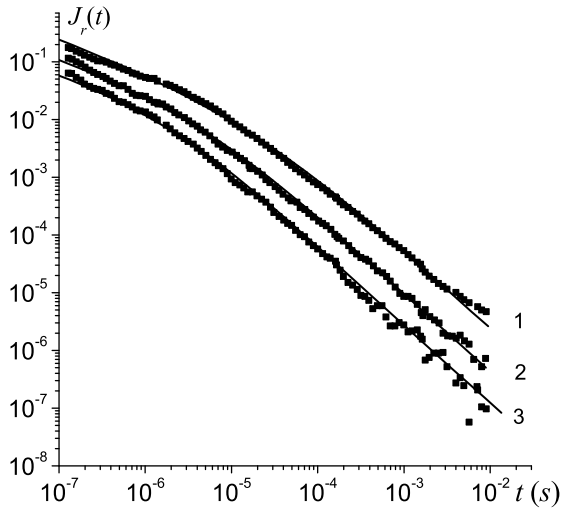}
\caption{Comparison of the time ($t$) dependence of SDAR flux $J_r
(t)$ measured experimentally\cite{SekiPRB} (black squares) and
calculated with eq. (\ref{sub15}) in the activated rate (or
kinetically controlled reaction) model of reactivity (full lines)
assuming weak reactivity, when $w_{e_s} \approx w_{e_a}$, and for:
(1) $T = 105 K, \, w_{_{0_s}} = 3.3 \cdot 10^4 s^{-1}, \gamma_a =
0.2$; (2) $T = 145 K, \, w_{_{0_s}} = 1.3 \cdot 10^5 s^{-1},
\gamma_a = 0.3$; (3) $T = 170 K, \, w_{_{0_s}} = 3.3 \cdot 10^5
s^{-1}, \gamma_a = 0.4$.}
\end{figure}

The fitting of theoretical $J_r (t)$-dependences to experimental
data (see Fig. 3) enables one to obtain corresponding parameters of
the activated rate model: $\alpha, \,w_{0_s}$ and $\gamma_a$.
Unfortunately, because of the very smooth shape of the kinetic
dependences  these parameters can be determined fairly
approximately. The results can be summarized as follows:

(1) The obtained values of $\alpha$ increase approximately linearly
with the increase of $T$ in agreement with results of earlier
analysis.\cite{SekiPRB} As for absolute values of the parameter
$\alpha$ they are about $10 - 15\%$ lower than those known for the
same system from time-of-flight experiments.\cite{SekiPRB} This
deviation, however, is quite within the accuracy of our method
(estimated to be $\sim 20\%$).

(2) The rate $w_{0_s} (T) $ increases with temperature $T$, as
expected: $w_{0_s}(105{\rm K}) \approx 3.3\cdot 10^4 \,
s^{-1},\:w_{0_s}(145{\rm K}) \approx  1.3\cdot 10^5 \, s^{-1},$ and
$w_{0_s}(170{\rm K}) \approx 3.3\cdot 10^{5}\, s^{-1}$.

(3)The parameter $\gamma_a $ is found to depend on temperature and
$\gamma_a (T)$ increases with $T$: $\gamma_a (105) \approx 0.2$,
$\gamma_a (145) \approx 0.3$, and $\gamma_a (170) \approx 0.4$ (the
estimated accuracy of these values is about $30\%$). This behavior
of $\gamma_a (T)$ can be qualitatively interpreted within the
proposed model. Under quite reasonable assumption of small reaction
rate in the well, $w_{r_a} \ll w_{e_s}$ when $w_{0_s} \approx
w_{e_s}$, one gets for the ratio
\begin{equation}\label{app6}
\gamma_a (T)/\gamma_a (T_0) \approx (T_0/T)
[w_{e_s}(T)/w_{e_s}(T_0)]^{\alpha}.
\end{equation}
In deriving this relation we took into account that $l_e (T)/l_e
(T_0) = T_0/T$ and assumed that $D_s$ depends on temperature weaker
than other (rate) parameters. In any case the temperature dependence
$\gamma_a (T)$ is mainly determined by that of $w_{e_s}(T)$.

Estimation [using eq. (\ref{app6})] yields: $\gamma_a (170)/\gamma_a
(105) \approx 1.40$ and $\gamma_a (145)/\gamma_a (105) \approx
1.31$. The corresponding values, obtained from fitting kinetic
curves, are given by: $\gamma_a (170)/\gamma_a (105) = 1.6 \pm 0.2$
and $\gamma_a (145)/\gamma_a (105) = 1.4 \pm 0.2$. Unfortunately low
accuracy of extracted parameters of the model does not allow
unambiguous test of predicted characteristic dependences.

Concluding this section it is worth discussing the results of our
analysis and comparing them with those obtained in the early work
[24] (see also ref. [23]), in which the kinetics of the same
process, geminate $e$-$h$ recombination in amorphous semiconductors
$a$-Si:H, is investigated both experimentally and theoretically in a
fairly wide region of temperatures: $8 < T < 150\, {\rm K}$.

One of principle differences of the interpretation, presented this
work, from our study lies in the applied model of migration of
recombining charges in the disordered semiconductors. In the work
[23] the analysis has been made within the Markovian model implying
the conventional Smoluchowski equation of type of eq. (\ref{gen1})
for the PDF, whereas in our study the migration is assumed to be
non-Markovian and is described with the use of the subdiffusion
model. The difference between these two descriptions, evidently,
shows itself in the long time behavior of the recombination
kinetics. Both models predict the inverse power type dependence of
the reaction yield on time $J_r(t) \sim 1/t^{1+\alpha/2}$, however
in the Markovian case $\alpha = 1$ while in the non-Markovian
(subdiffusion) one $\alpha < 1$. Analysis of recent experimental
measurements of the recombination kinetics\cite{SekiPRB} shows that
as applied to charge migration in amorphous $a$-Si:H semiconductors
the subdiffusion model is more realistic than the Markovian model.
This conclusion is strongly supported by recent time-of-flight
measurements of transient currents in these
semiconductors.\cite{Mur} Moreover, it is worth mentioning that
close inspection of experimental results presented in the work [24]
indicates some deviation of the long time part of the recombination
kinetics from the predicted dependence $1/t^{3/2}$. It is also seen
that at temperatures $T \gtrsim 60$ K the long time part is better
described by the function $ 1/t^{1+\alpha/2}$ with $\alpha < 1$.

The intermediate part of the luminescence decay (i.e. $e-h$
recombination) kinetics is described in ref. [24] by equation of
type of eq. (\ref{gen1}) with $u(r)=l_e/r$ and $k_r \sim
e^{-r/r_e}$. This complicated equation is solved approximately by
reducing it to the free diffusion one (which in turn is solved
within the prescribed diffusion approximation) and then evaluating
the effect of the Coulomb potential perturbatively. The weakness of
the effect of the interaction resulted from the assumed fairly
strong reactivity at short distances, i.e. fast reactive
disappearance of pairs which is expected to be insignificantly
affected by the potential. This assumption is, however, very
restrictive because in the process under study the Coulomb
interaction is quite strong: even at highest temperatures studied,
$T \approx 150 {\rm K}$, for realistic values of the contact
distance $d \approx 10 {\rm A}$ and dielectric constant $\varepsilon
\approx 10$ the dimensionless well depth is fairly large, $u(d)=
U(d)/(k_B T) \approx 10$. As for lower $T \approx 10 {\rm K}$, the
dimensionless well depth at these temperatures is even much ($15$
times) larger than that at $T \approx 150 {\rm K}$.

The observed reasonable accuracy of the perturbative solution of the
Smoluchowski equation is due assumed fairly strong and long distant
tunneling rate $k_e\sim \exp(-r/r_e)$, with the tunneling length
$r_e \approx 11 {\rm A}$. So large tunneling length corresponds to a
very small electron localization energy $E_l \approx 0.034 $ eV in
amorphous $a$-Si:H, which can easily be estimated by the relation
$r_e \approx r_B\sqrt{E_H/E_l}\,$,\cite{Lan} where $r_B \approx 0.5$
A and $E_H \approx 13.5$ eV are the Bohr radius and the ionization
energy of hydrogen atom, respectively. The localization energy
obtained appears to be very close to the thermal energy $E_{th}=
0.026$ eV corresponding to room temperature $T = 300$ K. Such a
small value of $E_l$ does not look quite realistic though some
arguments in favor of this estimate have been presented in ref.
[24].

In our interpretation, instead of assuming very small value of $E_l$
we properly described the effect of the well resulting from the
attractive Coulomb interaction and treated the reactivity suggesting
it to be weak enough to neglect its manifestation during the time of
population relaxation within the well. Because of long life time of
pairs in the well, however, this relatively weak reactivity strongly
manifests itself in the recombination kinetics leading to more
smooth recombination kinetics at intermediate times than that
predicted by free diffusion model in agreement with experimental
results.

The analysis and comparison with results of experiments at
relatively high temperatures $T \gtrsim 100$ K demonstrate that the
proposed subdiffusion variant of the two-state model makes it
possible to reproduce the behavior of the experimental luminescence
decay kinetics (i.e. $e$-$h$ recombination kinetics) both at
intermediate and long times thus describing fairly accurately the
kinetics in the whole region of experimentally investigated times.

Of certain interest is the observed luminescence decay kinetics at
very low temperature $T = 8$ K,\cite{Street1} the slope of which in
a semilogarithmic scale [i.e. the slope of the dependence $\ln
I(t)$] appeared to change non-monotonically with time at
intermediate times. In ref. [24] such a behavior is described
assuming the recombining pairs to be nearly immobile. In this case
the kinetics is determined by the specific features of $k_r \equiv
k_r (r)$ dependence and the initial DPF $\rho_i (r) = \rho (r,t=0)$,
which in ref. [24] is chosen to accurately fit the experimental
kinetics.

It is worth noting that in the proposed two-state model similar
non-monotonic behavior of the slope of $\ln I(t)$-dependence is
predicted in the first order reaction model (see Fig. 2). Such a
behavior found without special assumptions on the initial PDF.
However one has to assume that at low temperatures the first order
reaction mechanism of reactivity becomes more efficient than the
kinetically controlled reaction one (which seems to be of higher
efficiency at high temperatures). In the interpretation within the
first order reaction model one should also take into account
possible distribution of coordinates of the bottom of the well
which, for sure, will lead to the distribution of average
recombination rates. This distribution can somewhat modify the
kinetics as in the case of immobile reacting pairs.

\section{Conclusions}

This work concerns the analysis of the kinetics of geminate SDARs of
interacting particles. The interaction potential is assumed to be
attractive and of the well shape. The effect of this interaction is
shown to reduce to the formation of the quasistatic state within the
well (cage), which essentially controls the SDAR kinetics. The
reaction is suggested to occur only in the well. Two models of
reactivity in the well are discussed: the activated rate (or
kinetically controlled reaction) and the first order reaction
models. The results obtained have been used for the analysis of the
geminate electron-hole recombination in amorphous semiconductors
$a$-Si:H. This analysis have shown that the first (activated rate)
model is able to describe the experimental recombination kinetics
better than the second one and therefore can be considered as more
appropriate as applied to the processes in amorphous $a$-Si:H
semiconductors at not very low temperatures.

Comparison of experimental data with theoretical results have
enabled us to obtain the values of characteristic parameters of the
system under study within the proposed model. Unfortunately very
smooth experimental kinetic time dependences does not allow for
accurate enough determination of the parameters. In addition the
experimental results presented in refs. [19] and [24] cover only
relatively narrow temperature region in which parameters of the
model do not change strongly.

For the same reasons it is, strictly speaking, hardly possible to
make absolutely unambiguous conclusions in favor of any of two
above-mentioned reactivity models. The specific features of the
recombination kinetics at intermediate times predicted in the first
order reaction model, which seemingly disagree with those found
experimentally, can, nevertheless, be strongly smoothed in the
presence of the distribution of kinetic parameters of the two-state
model (average reaction rate $w_{r_f}$ and the escaping rate
$w_{e_s}$). Such a distribution is quite natural for studied
amorphous semiconductor.

The results of comparison show that more extensive experimental
investigations in a wider temperature range are desirable. These
studies would provide us with more detailed information on the
mechanisms of migration and interaction of quasiparticles in
amorphous semiconductors which would allow for making more reliable
conclusions about these mechanisms.

\textbf{Acknowledgements.}\, The author is grateful to Dr. V. P.
Sakun for valuable discussions. The work was partially supported by
the Russian Foundation for Basic Research.


\begin{thebibliography}{99}
\bibitem{Ca}
D. F. Calef and J. M. Deutch, Annu. Rev. Phys. Chem. {\bf 34}, 493
(1983).
\bibitem{Ri}
S. A. Rice, {\it Diffusion-limited reactions} (Elsevier, Amsterdam,
1985).
\bibitem{Ha}
P. H$\ddot {\rm a}$nngi, P. Talkner, and M. Bercovec, Rev. Mod.
Phys. {\bf 69}, 252 (1990).
\bibitem{Noo}
K. M. Hong and J. Noolandy, J. Chem. Phys. {\bf 68}, 5163 (1978); J.
Chem. Phys. {\bf 68}, 5172 (1978).
\bibitem{Shu1}
A. I. Shushin, Chem. Phys. Lett. {\bf 118}, 197 (1985).
\bibitem{Shu2}
A. I. Shushin, J. Chem. Phys. {\bf 95}, 3657 (1991).
\bibitem{Shu3}
A. I. Shushin, J. Chem. Phys. {\bf 97}, 1954 (1992).

\bibitem{Met1} R. Metzler and J. Klafter, Phys. Rep. \textbf{339}, 1 (2000).

\bibitem{Blum}
A. Blumen, J. Klafter and G. Zumofen, {\it in} Fractals in physics,
edited by L. Pietronero and E. Tosatti (North Holland, Amsterdam,
1986), p. 399.

\bibitem{Yuste}
S. B. Yuste and K. Lindenberg, Phys. Rev. Lett. {\bf 87}, 118301
(2001); S. B. Yuste and K. Lindenberg, Chem. Phys. {\bf 284}, 169
(2002).

\bibitem{Yuste2004}
S. B. Yuste, L. Acedo and K. Lindenberg, Phys. Rev. E {\bf 69},
036126 (2004).

\bibitem{Henry}
B. I. Henry and S. L. Wearne, Physica A {\bf 276}, 448 (2000); B. I.
Henry and S. L. Wearne, SIAM J. Appl. Math. {\bf 62}, 870 (2002).

\bibitem{Vlad}
M. O. Vlad and J. Ross, Phys. Rev. E {\bf 66}, 061908 (2002).

\bibitem{Fed}
S. Fedotov and V. M\'{e}ndez, Phys. Rev. E {\bf 66}, 030102 (2002).

\bibitem{Sung}
J. Sung, E. Barkai, R. J. Silbey, and S. Lee, J. Chem. Phys. {\bf
116}, 2338 (2002).

\bibitem{ShushinPRE}
A. I. Shushin, Phys. Rev. E {\bf 67}, 061107 (2003).

\bibitem{Shushin}
A. I. Shushin, J. Chem. Phys. {\bf 122}, 154504 (2005).

\bibitem{ShushinNJ}
A. I. Shushin, New J. Phys. {\bf 7}, 21 (2005).

\bibitem{Seki}
K. Seki, M. Wojcik, and M. Tachiya, J. Chem. Phys. {\bf 119}, 2165
(2003); J. Chem. Phys. {\bf 119}, 7525 (2003); J. Chem. Phys.  {\bf
124}, 044702 (2006).


\bibitem{Sokolov06} I. M. Sokolov, M. G. W. Schmidt, and F. Sagu\'{e}s,
Phys. Rev. E {\bf 73}, 031102 (2006).

\bibitem{Kenkre}
V. M. Kenkre, E. W. Montroll and M. F. Shlesinger, J. Stat. Phys.
{\bf 9}, 45 (1973); E. W. Montroll and M. F. Shlesinger, {\it
Studies of Statistical Mechanics}, edited by J. L. Lebowitz and E.
W. Montroll (North Holland, Amsterdam, 1984), vol. 11, p. 5.

\bibitem{Met2} E. Barkai, R. Metzler, and J. Klafter, Phys. Rev. E \textbf{61}, 132 (2000).

\bibitem{Street}
R. A. Street, {\it Adv. Phys.} {\bf 30}, 593 (1981).

\bibitem{Street1}
K.M. Hong, J. Noolandy, and R. A. Street, Phys. Rev. B {\bf 23},
2967 (1981).

\bibitem{SekiPRB} K. Seki, K. Murayama, and M. Tachiya, Phys. Rev. B
{\bf 71}, 235212 (2005).

\bibitem{Shu4} A. I. Shushin, Khim. Fiz. (Russ. Chem. Phys.) {\bf 1}, 1217
(1982); S. H. Robertson, A. I. Shushin, and D. M. Wardlaw, J. Chem.
Phys. {\bf 98}, 8673 (1993),

\bibitem{Shu5} A. I. Shushin and E. Pollak, J. Chem. Phys. {\bf 98}, 8673 (1993).

\bibitem{Samko}
S. G. Samko, A. A. Kilbas, O. I. Marichev {\it Fractional Integrals
and Derivatives - Theory and Applications} (Gordon and Breach, New
York, 1993).

\bibitem{Erd}
A. Erd\'elyi, {\it Tables of Integral Transforms, Bateman Manuscript
Project}, Vol. I (McGraw-Hill, New York, 1954).

\bibitem{Nin} D. C. J. Chan, D. J. Mitchell, and B. W. Ninham,
J. Chem. Phys. {\bf 70}, 2946 (1979).

\bibitem{Ovch}
A. A. Ovchinnikov, S. F. Timashev, and Belyi {\it Kinetics of
Diffusion-Controlled Chemical Processes} (Chemistry, Moscow, 1986).

\bibitem{Mur} K. Murayama, and Y. Ando, Phys. Stat. Solidi C
{\bf 1}, 117 (2004).

\bibitem{Lan}
L. D. Landau and E. M. Lifshitz {\it Quantum Mechanics} (Pergamon,
Oxford, 1965).


\end{thebibliography}
\end{document}